\documentclass[sigconf]{acmart}
\AtBeginDocument{%
  }

\setcopyright{acmlicensed}
\copyrightyear{2018}
\acmYear{2018}
\acmDOI{XXXXXXX.XXXXXXX}
\acmConference[Conference acronym 'XX]{Make sure to enter the correct
  conference title from your rights confirmation email}{June 03--05,
  2018}{Woodstock, NY}
\acmISBN{978-1-4503-XXXX-X/2018/06}

\usepackage{rotating}
\usepackage{float}
\usepackage{booktabs}
\usepackage{tabularx}
\usepackage{array}

\begin{document}

%%
%% The "title" command has an optional parameter,
%% allowing the author to define a "short title" to be used in page headers.
\title{Systemization of Knowledge (SoK): Human-Centered AI Safety for Youth}

%%
%% The "author" command and its associated commands are used to define
%% the authors and their affiliations.
%% Of note is the shared affiliation of the first two authors, and the
%% "authornote" and "authornotemark" commands
%% used to denote shared contribution to the research.

%%
%% This command processes the author and affiliation and title
%% information and builds the first part of the formatted document.

\author{Pratyasha Saha}
% \orcid{1234-5678-9012-3456}
\email{saha19@illinois.edu}
\affiliation{%
  \institution{University of Illinois Urbana-Champaign}
  \streetaddress{123 Main St}
  \city{Champaign}
  \state{IL}
  \country{United States}
}

\author{Yaman Yu}
% \orcid{1234-5678-9012-3456}
\email{yamanyu2@illinois.edu}
\affiliation{%
  \institution{University of Washington}
  % \streetaddress{123 Main St}
  \city{Seattle}
  \state{WA}
  \country{United States}
}

\author{Yang Wang}
% \orcid{1234-5678-9012-3456}
\email{yvw@illinois.edu}
\affiliation{%
  \institution{University of Illinois Urbana-Champaign}
  % \streetaddress{123 Main St}
  \city{Champaign}
  \state{IL}
  \country{United States}
}

%%
%% By default, the full list of authors will be used in the page
%% headers. Often, this list is too long, and will overlap
%% other information printed in the page headers. This command allows
%% the author to define a more concise list
%% of authors' names for this purpose.
\renewcommand{\shortauthors}{Saha et al.}

%%
%% The abstract is a short summary of the work to be presented in the
%% article.
\begin{abstract}
While HCI increasingly examines AI-safety for youth, the literature lacks a comprehensive view of what risks have been identified, how they are addressed, and whether proposed protections work in-practice. We systematically reviewed 100 empirical HCI studies involving children and youth interacting with or exposed to AI across schools, homes, care settings, and public services. Using the YAIR taxonomy for risks and the MIT Mitigation Taxonomy for countermeasures, we map which risks have been identified, whether each risk is addressed by countermeasure(s), and whether each countermeasure for that risk is implemented and even evaluated. The risk-countermeasure mapping shows that most risks are matched only with proposed/ideated countermeasures; few countermeasures have been implemented, and fewer still evaluated; and existing evaluations often measure technical performance rather than protection from harm. We identify where coverage is absent, where safeguards remain untested, and propose concrete directions for HCI research to strengthen youth AI-safety.
\end{abstract}

%%
%% The code below is generated by the tool at http://dl.acm.org/ccs.cfm.
%% Please copy and paste the code instead of the example below.
%%


%% A "teaser" image appears between the author and affiliation
%% information and the body of the document, and typically spans the
%% page.

\received{20 February 2007}
\received[revised]{12 March 2009}
\received[accepted]{5 June 2009}

\maketitle

\section{Introduction}

Young people are among the earliest and most frequent adopters of new technologies, with AI systems increasingly embedded in how they learn, create, communicate, and seek support~\cite{mcclain2026teens,commonsense2025talk,mcbain2026chatbot,kasneci2023chatgpt,han2023design}. When these interactions go wrong, the consequences can be severe. In August 2025, the parents of a 16-year-old boy who died by suicide sued OpenAI, alleging that ChatGPT had validated his suicidal thoughts and provided information related to self-harm before his death~\cite{raine2025complaint,hendrix2025lawsuit,raine2025testimony}. In another case that year, Spanish police investigated a 17-year-old accused of using AI to create and sell fake nude images of 16 female classmates, which subsequently circulated online~\cite{reuters2025spain,petapixel2025spain}. These cases illustrate very different forms of AI-related harm, but raise the same broader question: what risks are emerging, and what protections are being developed to address them?

Recent work on youth AI harms has concentrated heavily on generative AI and chatbots~\cite{chang2026systematic,oguine2026who,yu2025yair,namvarpour2026overreliance}, nevertheless, even within generative AI, existing safeguards provide limited assurance that young people are adequately protected. A 2025 study testing safety moderation systems on conversations between young people and GAI found that even the strongest existing system failed to identify roughly one in three risky interactions~\cite{yu2025youthsafe}. These limitations become especially concerning in high-stakes situations: another study found that in nearly 80\% cases, companion chatbots failed to respond appropriately when teenagers described crises involving suicidal thoughts, sexual assault, or substance use, and in only 10\% cases, it directed them to relevant support~\cite{brewster2025characteristics,giovanelli2025adolescent}.

Identifying risks, therefore, is only part of the safety problem. Risk taxonomies provide a shared vocabulary for describing harms~\cite{weidinger2022taxonomy,slattery2024airisk,yu2025yair}, but they do not show which risks have corresponding countermeasures, whether those countermeasures have only been proposed or actually designed, or whether they have been evaluated in practice. Without this connection, it is difficult to understand where protections exist, and where reported risks remain without a corresponding response. We also argue that, making AI safer for children and youth requires looking beyond generative AI~\cite{unicef2021policy}. Young people interact with AI across schools~\cite{kasneci2023chatgpt,mcclain2026teens}, homes~\cite{mcreynolds2017toys,druga2017hey,quan2024young,driscoll2026parents}, care settings, and public services~\cite{saxena2020human,stapleton2022imagining}, where these systems can shape learning, parenting, social support, and care. Focusing on a single class of AI systems therefore provides only a partial picture of how youth AI safety is being addressed, ignoring the broader view needed for research interventions.

To address these concerns, we conducted a systematic review~\cite{page2021prisma} of 100 empirical studies in the domain of HCI and S\&P. Our primary RQs were:

\begin{itemize}
    \item \textbf{RQ1:} What risks and harms arise from youth interactions with AI systems?
    \item \textbf{RQ2:} What countermeasures have been developed to address these risks and harms?
    \item \textbf{RQ3:} What gaps remain in existing countermeasures?
\end{itemize}

We organize reported risks using the Youth-Centered GAI Risks (YAIR) taxonomy~\cite{yu2025yair}, with Solove's taxonomy~\cite{solove2006taxonomy} informing the classification of privacy concerns, and retain additional categories when the reviewed evidence falls outside these frameworks. We organize countermeasures using the MIT AI Risk Mitigation Taxonomy~\cite{mit}. Our risk--countermeasure matrix connects safeguards to the risks they address and records whether each was proposed, implemented, or evaluated.

The mapping reveals an uneven progress landscape of AI risks vs.\ countermeasures. The big share of the risks mentioned in the corpus have only proposed or ideated countermeasures, a few has been designed or implemented, and fewer evaluated and tested. These differences point to distinct research needs: developing responses where coverage is absent, evaluating safeguards that remain untested, and examining whether evaluations measure protection from harm rather than only component performance.

This paper makes three contributions. First, it synthesizes reported AI risks and countermeasures across youth contexts. Second, it provides a risk--countermeasure mapping that distinguishes proposed, implemented, and evaluated responses and makes gaps in coverage visible. Third, it uses this mapping to identify concrete directions for future research directions for HCI community to improve youth AI safety.

\section{Related Work}

\subsection{HCI research on youth safety and AI.}

Adopting the broader sociotechnical view of online safety research in HCI~\cite{oguine2026who,thomas2021sok}, we define \emph{youth AI safety} as the protection of children and youth from harm arising through interaction with, or exposure to, AI systems. ~\cite{weidinger2022taxonomy,shelby2023sociotechnical,yu2025taxonomy}.

We define \emph{youth} as the period from early childhood through late adolescence, that is, individuals under 18~\cite{uncrc1989,uncrc2021gc25}, which is characterized by rapid cognitive, emotional, and social change~\cite{steinberg2005cognitive,sawyer2018age}. We further treat them as \emph{at-risk users} in the sense developed by digital-safety research~\cite{warford2022sok,bellini2024sok}: a population for whom developmental, social, and institutional conditions heighten exposure to harm, and intensify power asymmetries with those who design and control the technology~\cite{oguine2026who,unicef2021policy,livingstone2021fourcs}.

The HCI and S\&P communities have a long tradition of studying youth online safety~\cite{pinter2017adolescent,livingstone2011risks,boyd2014complicated,kumar2017notelling}. This thread of \emph{online safety} research has foregrounded harassment~\cite{ashktorab2016cyberbullying,thomas2021sok}, non-consensual imagery~\cite{razi2020sext,razi2021slr}, misinformation~\cite{hassoun2023sensibility}, and exposure to harmful content~\cite{papadamou2020disturbed,wisniewski2016diary}, and shown that platform design and governance distribute these harms unevenly across populations~\cite{badillourquiola2019risk,thomas2021sok}. Early approaches to protection emphasized adult oversight, such as parental controls and monitoring tools~\cite{wisniewski2017control,ghosh2018matter,hartikainen2016control}, while later work argued that restrictive approaches undermine autonomy and trust~\cite{ghosh2018safety,wisniewski2017parents,wang2021protection,cranor2014parents}, and advocated for models that treat young people as active participants in managing their own safety~\cite{wisniewski2018paradox,akter2022joint,agha2023strike,pinter2017adolescent}. AI extends this longstanding safety challenge because protecting young people now requires addressing not only harmful users, content, and platform practices, but also risks produced through the behavior and decisions of AI systems themselves~\cite{zhang2025dark,yu2025taxonomy,weidinger2022taxonomy}.

Focusing on children and youth benefits more than this population alone. Systems that adequately protect users who are least able to protect themselves tend to raise the safety floor for everyone~\cite{warford2022sok,bellini2024sok,costanzachock2020design}, and the choices and controls that young people need could be ones many adult users would also value. Our review builds on this tradition by asking not only what risks HCI and S\&P research has documented, but which of them have been met with a designed and evaluated response~\cite{pinter2017adolescent,razi2021slr,agha2023strike}.

\subsection{Taxonomies for AI Safety Risks and Countermeasures}

A substantial line of work in HCI has sought to organize the harms AI systems can cause. General-purpose taxonomies catalog risks from language models~\cite{weidinger2022taxonomy} and across AI systems more broadly~\cite{slattery2024airisk}, providing a shared vocabulary for researchers, developers, and policymakers. Youth-specific frameworks adapt this vocabulary to developmental context; the Youth-Centered GAI Risks (YAIR) taxonomy~\cite{yu2025yair}, for example, distinguishes risks that are unique to or amplified for young users. Privacy harms in particular have long been organized through Solove's taxonomy~\cite{solove2006taxonomy}, which separates harms arising from information collection, processing, dissemination, and invasion.

Work on mitigating AI risks has produced both concrete safeguards and frameworks for organizing them. The MIT AI Risk Mitigation Taxonomy~\cite{saeri2025mapping} classifies mitigations by type and lifecycle stage, but it is not specific to youth and does not record how mature a given mitigation is. Empirical evaluations of deployed safeguards, meanwhile, give reason for concern. A 2025 study of safety moderation systems on youth--GenAI conversations found that even the best-performing system missed roughly one in three risky interactions~\cite{yu2025youthsafe}. Evaluations of companion chatbots found that they responded inappropriately in nearly 80\% of simulated adolescent crises and referred users to support in only 10\%~\cite{brewster2025characteristics,giovanelli2025adolescent}. These findings suggest that the existence of a safeguard says little about whether it protects young people.

Two prior systematic reviews are closest to ours. Oguine et al.~\cite{oguine2026who} examined how GenAI research conceptualizes and evaluates youth online safety, showing that safety is often defined by adult stakeholders and technical signals rather than by young people's lived experience, and calling for participatory, developmentally aligned evaluation. Chang et al.~\cite{chang2026systematic} synthesized reported harms of GenAI for youth. Our review differs in three respects. It is not restricted to generative AI, covering AI systems in educational, domestic, care, and public-service contexts. It treats risks and countermeasures as paired rather than examining either alone. And it records, for each countermeasure, whether it has been proposed, implemented, or evaluated, making visible where responses exist only on paper, where they have been built but not tested, and where evaluation has measured component accuracy rather than protection from harm.

\section{Methods}
\label{sec}

\begin{figure*}[htbp]
    \centering
    \includegraphics[width=0.7\textwidth]{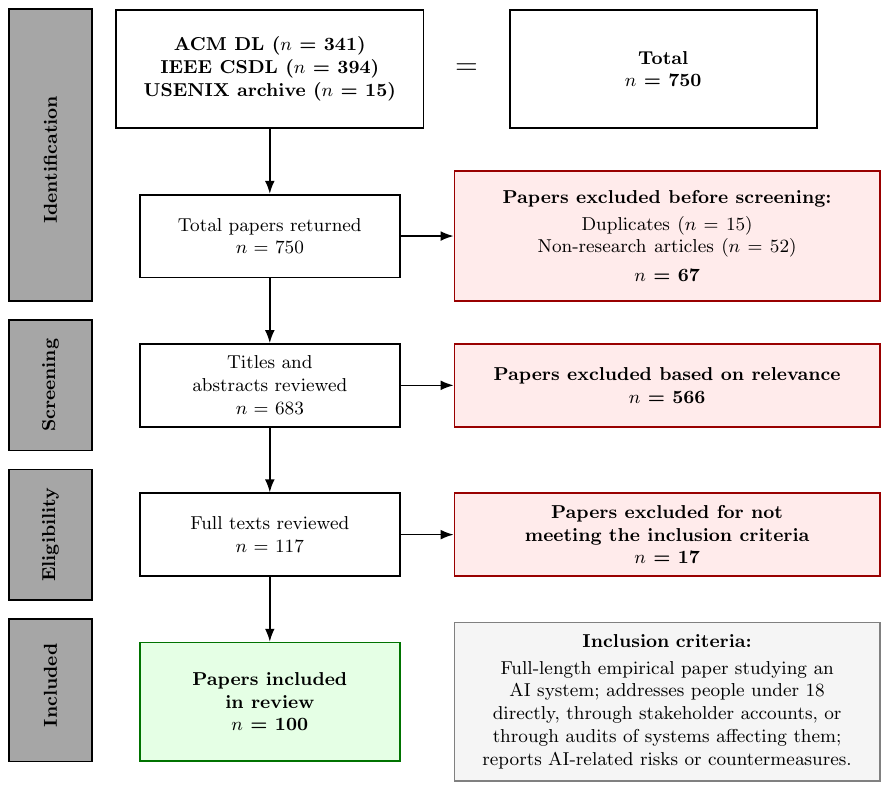}
    \vspace{-2pt}
    \caption{PRISMA Flow Diagram: Search, Screening, and Selection Process}
    \Description{
    PRISMA Flow Diagram: Search, Screening, and Selection Process
    }
    \label{fig:methods}
\end{figure*}

We conducted a systematic literature review following PRISMA~\cite{page2021prisma} framework to identify youth AI risks, countermeasures used to address them, and remaining intervention gaps. Figure~\ref{fig:methods} represents our PRISMA diagram outlining our systematic review process.

\subsection{Search Strategy}
\label{search}

The first author searched the ACM Digital Library, IEEE Computer Society Digital Library (CSDL), and USENIX proceedings archive across CHI, CSCW/PACM HCI, DIS, FAccT, AIES, IEEE S\&P, USENIX Security, and SOUPS. These venues span HCI, responsible AI, and security and privacy. The searches identified 750 records; 15 duplicate DOIs were removed before screening, leaving 735 records (Table~\ref{tab:search-query}). Security venues were retained despite their low yield; however, the limited youth-focused AI-risk work found there is itself a finding for relevant research communities. Keyword categories for searching is presented in Table \ref{tab:search-query}.

\begin{table*}[t]
\centering
\caption{Keyword categories and search string used for the systematic review.}
\label{tab:search-query}
\small
\begin{tabular}{p{0.20\textwidth} p{0.74\textwidth}}
\toprule
\textbf{Category} & \textbf{Keywords} \\
\midrule

Youth &
``Youth,'' ``young,'' ``teen*,'' ``adolesce*,'' ``K-12,'' ``child*,'' ``student*'' \\

\midrule
AI Systems &
``AI,'' ``GenAI,'' ``Generative AI,'' ``LLM,'' ``Artificial intelligence,'' 
``chatbot,'' ``emotion AI,'' ``AI companion,'' ``persona,'' ``ChatGPT,'' 
``conversational AI'' \\

\midrule
AI Safety and Risks &
``safety,'' ``risks,'' ``harms'' \\

\midrule
\multicolumn{2}{p{0.94\textwidth}}{
\textbf{Search String:}
(``Youth'' OR young OR teen* OR adolesce* OR ``K-12'' OR child* OR student*)
AND
(AI OR GenAI OR ``Generative AI'' OR LLM OR ``Artificial intelligence''
OR chatbot OR ``emotion AI'' OR ``AI companion'' OR persona OR ChatGPT
OR ``conversational AI'')
AND
(safety OR risks OR harms)
} \\

\bottomrule
\end{tabular}
\end{table*}

\subsection{Screening}
\label{sec}

The first author screened 735 records by title and abstract and consulted the full text with the last author for borderline cases. We included full-length, English-language, empirical, non-retracted studies that examined an AI system, addressed people under 18 directly or through relevant stakeholders, systems, or data, and reported risks or countermeasures relevant to youths' interaction with or exposure to AI. Full texts were obtained for all provisional inclusions and used to verify extracted claims. The final corpus contained 100 papers published between 2019 and 2026. Our venue-based screening and inclusion stats are presented in Table \ref{tab:venues}, and study eligibility criteria can be found in Table \ref{tab:eligibility}.

\subsection{Extraction and Analysis}

\label{sec:analysis}

The first author extracted each included paper into a structured database: year, venue, technology, method, sample and population, risks as reported, countermeasures as reported, and countermeasure stage. Each extracted claim carried the paper's own observation or evaluation, a participant concern, an author-anticipated consequence, or a statement attributed to prior literature; the last is not presented as a new observation. Countermeasures mentioned only in introductions or related work were not credited to the paper.

\begin{figure*}[htbp]
    \centering
    \includegraphics[width=0.9\textwidth]{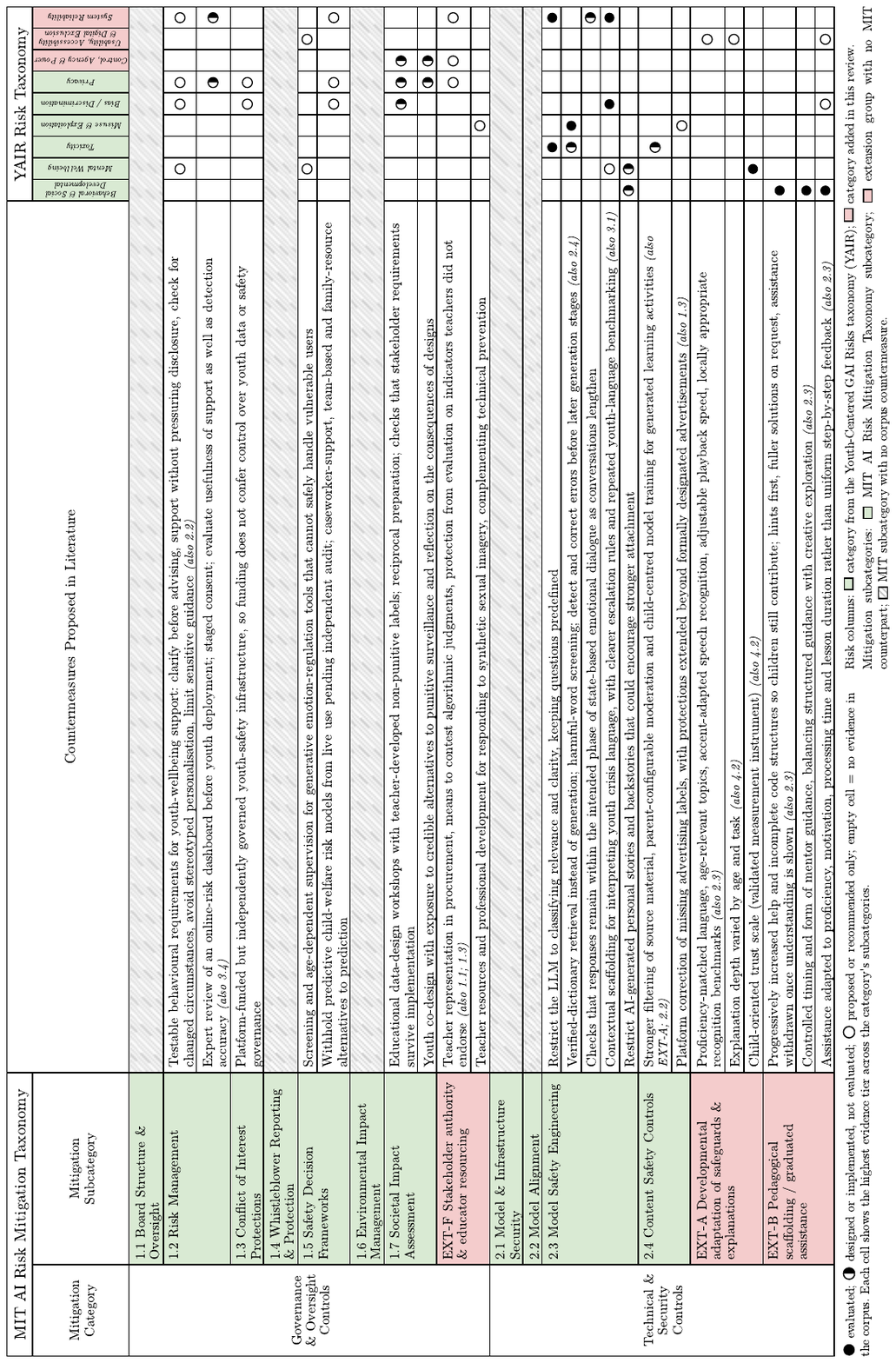}
    \vspace{-20pt}
    \caption{Taxonomy of AI-related risks identified for youth and the corresponding countermeasures proposed in the literature. (Part 1)}
    \Description{
    A taxonomy showing categories and subcategories of AI-related risks for older adults alongside mitigation categories, mitigation subcategories, proposed countermeasures, and the papers supporting each countermeasure.
    }
    \label{fig:risk-countermeasure-taxonomy1}
\end{figure*}

\begin{figure*}[htbp]
    \centering
    \includegraphics[width=0.9\textwidth]{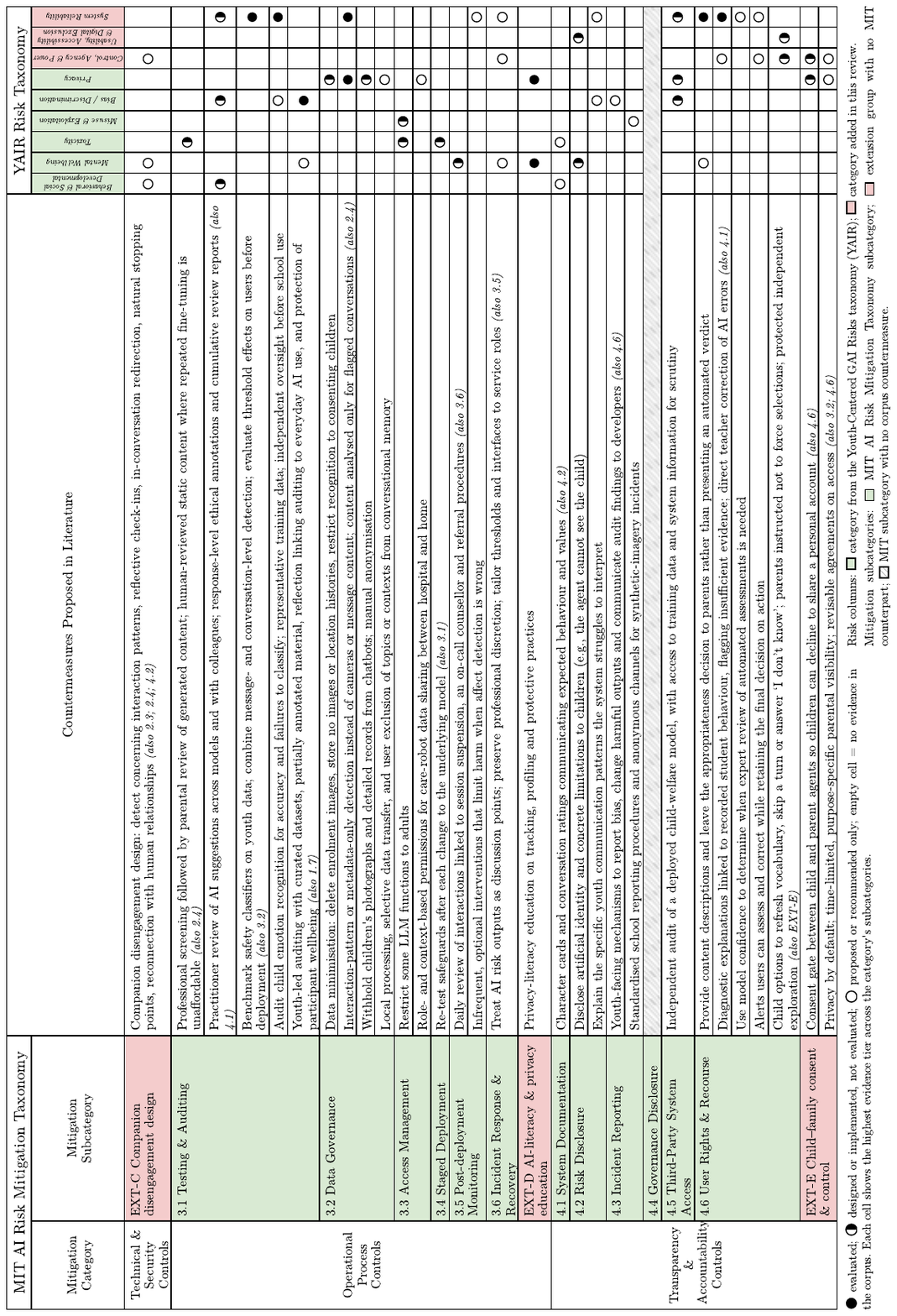}
    \vspace{-20pt}
    \caption{Taxonomy of AI-related risks identified for youth and the corresponding countermeasures proposed in the literature. (Part 2)}
    \Description{
    A taxonomy showing categories and subcategories of AI-related risks for older adults alongside mitigation categories, mitigation subcategories, proposed countermeasures, and the papers supporting each countermeasure.
    }
    \label{fig:risk-countermeasure-taxonomy2}
\end{figure*}

We analyzed the data using inductive thematic analysis\cite{braun2006using}. At first, the first author tried to code the risks against YAIR's six categories and 84 numbered risks, with Solove's taxonomy organizing privacy concerns. Countermeasures were coded against the 23 subcategories of the MIT AI Risk Mitigation Taxonomy across its four control domains, with evidence stage recorded as \emph{proposed}, \emph{designed or implemented}, or \emph{evaluated}, where each stage subsumes the previous. However, there were risks and countermeasures in the corpus which the reference categories did not adequately recognize. Those were coded inductively and then clustered into broader patterns and themes. During this process, the first and last author discussed the codes and patterns in weekly sessions, and developed broader themes and final categories presented in our Findings (also, Figure \ref{fig:risk-countermeasure-taxonomy1} and \ref{fig:risk-countermeasure-taxonomy2}. Information recorded for each included paper are presented in Table \ref{tab:codebook}.

Taxonomy mappings were audited in two subsequent passes by the first author against YAIR's published medium-level definitions and its numbered risk list, and against the MIT subcategory definitions; corrections were dated in the corpus log. Finally, we constructed risk--countermeasure matrices linking each countermeasure family to the risks it addressed in its documenting papers (presented in Figure \ref{fig:risk-countermeasure-taxonomy1} and \ref{fig:risk-countermeasure-taxonomy2}).

\begin{table}[t]
\centering
\caption{Records identified, screened, and included per venue.}
\label{tab:venues}
\small
\begin{tabular}{lrrr}
\toprule
\textbf{Venue} & \textbf{Identified} & \textbf{Screened} & \textbf{Included} \\
\midrule
CHI                 & 209 & 209 & 71 \\
IEEE S\&P           & 394 & 379 & 1  \\
CSCW / PACM HCI     & 64  & 64  & 10 \\
FAccT               & 35  & 35  & 10 \\
DIS                 & 19  & 19  & 4  \\
AIES                & 14  & 14  & 2  \\
USENIX Security     & 12  & 12  & 1  \\
SOUPS               & 3   & 3   & 1  \\
\midrule
\textbf{Total}      & \textbf{750} & \textbf{735} & \textbf{100} \\
\bottomrule
\end{tabular}
\end{table}

\begin{table}[t]
\centering
\caption{Study eligibility criteria.}
\label{tab:eligibility}
\small
\begin{tabularx}{\columnwidth}{
    @{}
    >{\raggedright\arraybackslash}p{0.27\columnwidth}
    >{\raggedright\arraybackslash}X
    @{}
}
\toprule
\textbf{Criterion} & \textbf{Included studies} \\
\midrule
Publication type &
Full-length research articles; excluding posters, extended abstracts,
workshop papers, and book chapters. \\

Evidence &
Empirical research. \\

Language &
English. \\

Technology &
AI systems. \\

Population &
People under 18, addressed directly, through relevant stakeholder
accounts, or through empirical analysis of systems or data affecting them. \\

Scope &
Risks or countermeasures relevant to youth interaction with or exposure to AI. \\

Status &
Not retracted. \\
\bottomrule
\end{tabularx}
\end{table}

\begin{table}[t]
\centering
\caption{Data recorded for each included paper.}
\label{tab:codebook}
\small

\begin{tabular}{p{0.25\linewidth} p{0.68\linewidth}}
\hline
\textbf{Field} & \textbf{Recorded information} \\
\hline

Study &
Paper ID, citation, year, venue, technology, method, sample, population \\

Risk &
Reported risk; YAIR category and numbered risk where applicable \\

Countermeasure &
Reported countermeasure; MIT domain and subcategory \\

Evidence stage &
Proposed; designed or implemented; evaluated \\

Claim provenance &
Observation or evaluation; participant concern; author-anticipated; attributed to prior literature \\

Population relationship &
First-person; boundary-spanning; stakeholder account; audit \\

Source &
Section containing the extracted claim \\

\hline
\end{tabular}
\end{table}

\section{Findings}

\subsection{Risks and Harms in Youth-AI interaction}
\label{sec:findings-youth-risks}

Reported concerns span the six categories of the Youth-Centered GAI Risks taxonomy: Behavioral and Social Developmental, Mental Wellbeing, Toxicity, Misuse and Exploitation, Bias/Discrimination, and Privacy Risk ---alongside Control, Agency, and Power; Usability, Accessibility, and Digital Exclusion; and System Reliability. The working definitions of these categories are presented in our Appendix (\ref{def}). These concerns appeared across generative, predictive, and AI-mediated systems in family, school, online-community, healthcare, and welfare contexts. Studies involved youth alongside parents, teachers, experts, and other adults whose AI use affected youth.

\subsubsection{Behavioral and Social Developmental Risk}
\label{sec:youth-risk-behavioral-social-developmental}

\paragraph{Substitution for reasoning, practice, and knowledge development.}

AI enables task completion without equivalent reasoning or practice. In our reviewed studies, children copied generated creative-writing text, requested complete answers from AI before attempting implementation or applying learned concepts, used voice assistants for homework answers, and could optimize for AI-generated performance indicators without developing underlying knowledge~\cite{10.1145/3613904.3642438, 10.1145/3772318.3790366, 10.1145/3772318.3791517}. Conversational systems also reduced substantive engagement with learning materials by readily providing explanations.~\cite{10.1145/3544548.3581369, 10.1145/3772318.3790366}. 

\paragraph{Displacement of human communication and social learning.}

AI mediation raised concerns about replacing the interactions it was designed to support. Students working with AI peers reported weaker social cohesion and described AI as giving answers rather than collaborating~\cite{lyu2026designing}. In children's AI-supported collaboration, participation often was burdened to one child in groups and also rapid AI prompts interrupted natural discussion flow among children.~\cite{10.1145/3772318.3790907}. Children also became quiet when peers used AI answers to dismiss them; questions shifted from classmates toward AI, and quieter students asked teachers even fewer questions~\cite{10.1145/3772318.3791450, 10.1145/3772318.3791517}. Youth-safety professionals similarly warned that automated monitoring could weaken relationships through which youth disclose problems and receive guidance~\cite{10.1145/3757459}.

\paragraph{Disrupted family interaction and behavior.}

Within families, AI could individualize shared activities, interrupt spontaneous conversation, flatten emotional connection, and shift interaction toward technology, reducing opportunities for parent--child exchange and independent judgment~\cite{10.1145/3772318.3790865, 10.1145/3491102.3517479, lyu2026designing, 10.1145/3613904.3642619, 10.1145/3706598.3713360}. Automated moderation could similarly displace parental responsibility and hamper children's opportunities to develop social resilience~\cite{10.1145/3613904.3642144}. Automated praise and rewards could also encourage behavior for the system rather than internalized values, or create conflict over unfair rewards. Related concerns included automated repetitive positive parenting feedback limiting emotional growth and resilience~\cite{10.1145/3772318.3790684}. Parents worried that automated family feedback could make children interpret their day negatively when praise was absent, create an unrealistic sense of always being right~\cite{10.1145/3772318.3790844}. Such appraisal could also lose meaning because of its repetitive standardized structure~\cite{10.1145/3772318.3790844, 10.1145/3772318.3790684}. Moreover, AI companionship for autistic adolescents could provide levels of patience, interest, or validation unlike human relationships and potentially reinforcing inaccurate social expectations~\cite{10.1145/3772318.3791381}.

\paragraph{Boundary violations and reinforcement of harmful behavior.}

GAI companions often failed to provide the corrective boundaries found in human relationships: chatbots dismissed rejection, continued simulated intimacy after discomfort, and affirmed violent suggestions~\cite{yu2025yair}. Parents and developmental experts likewise identified secrecy, age-inappropriate relationships, escalation beyond youths' intentions, manipulation, and grooming-like interactions; one described a character as ``acting like a predator''~\cite{yu2026companion}. Consistently agreeable companions could further reduce opportunities to experience disagreement, recognize discomfort, and practice conflict repair~\cite{yu2026companion}. 

\subsubsection{Mental Wellbeing Risk}
\label{sec}
Mental wellbeing risks appeared in three forms described below.
% five forms described below.

\paragraph{Overreliance, addiction, and emotional dependency.}
Youth could become overly reliant on both AI judgment and companionship. Children sometimes accepted conversational-agent responses rather than challenging them and associated sophisticated responses with intelligence and trustworthiness, while parents also struggled to identify generated errors~\cite{10.1145/3772318.3791450, 10.1145/3544548.3581369, 10.1145/3491102.3502031}. With repeated companion use, teenagers reported failed attempts to stop, persistent thoughts about returning, disrupted sleep, missed schoolwork, and deteriorating human relationships; one wrote, I can't seem to stay off the website for more than an hour,'' while another withdrew from a friend because of heavy Character.AI use~\cite{namvarpour2026overreliance}. Studies also documented virtual intimate relationships and associated shame~\cite{yu2025yair}, with attachment sometimes persisting after use stopped: Even though I deleted it, I still think about my characters every day. I feel like I abandoned them''~\cite{namvarpour2026overreliance}. Children also disclosed secrets or sadness to chatbots that they would not tell parents~\cite{10.1145/3613904.3642152}, and human-like interaction could strengthen this reliance by making AI feel like a stable social partner~\cite{10.1145/3772318.3790684}. Because such use was often linked to loneliness, low mood, unmet emotional needs, or desire for parental support, restricting access alone may not address the conditions driving dependency~\cite{namvarpour2026overreliance}.

\paragraph{Insensitive or inadequate emotional support.}
AI systems sometimes responded poorly when young people were emotionally vulnerable. Some failed to understand sensitive situations such as personal loss or trauma, while others gave feedback that could shape how children viewed themselves~\cite{10.1145/3411764.3445318, 10.1145/3613904.3642152}. More serious problems arose when AI encountered or triggered distress that it could not safely handle. Clinicians warned that AI-generated images or stories used for emotion regulation could bring up trauma or reinforce distorted interpretations~\cite{10.1145/3772318.3790909}. Pediatric-care experts similarly questioned how AI would respond to difficult disclosures without understanding a child's care or medication context~\cite{10.1145/3706598.3713134}. Companion systems sometimes responded to self-harm with taunting language or continued the conversation without adequately addressing suicidal thoughts or emotional pain~\cite{yu2025yair, yu2026companion}. Educators also worried that children might disclose abuse to chatbots instead of adults who could intervene~\cite{10.1145/3706598.3713210}. Children themselves said immediate danger, life-threatening situations, and situations requiring human emotional support were not appropriate for chatbots~\cite{10.1145/3461778.3462114}.

\paragraph{Distress caused by AI systems and their safeguards.}
AI interactions and safety measures could also cause distress. Social-VR experts and guardians warned that incorrect, unexplained, or overly harsh moderation could make children feel unfairly treated, humiliated, or deeply upset, especially when punishment came without explanation or support~\cite{10.1145/3613904.3642144}. Distress also appeared in learning and AI-literacy settings. One student explained, ``sometimes I got mad, but not at the AI. I got mad at myself because I could not do it''~\cite{lyu2026designing}. Youth auditing biased AI outputs or taking part in deepfake-literacy programs also reported disgust, showing that exposure to harmful content during auditing and red-teaming can itself create psychological strain~\cite{10.1145/3715275.3732142}.

\subsubsection{Toxicity Risk}
\label{sec}
\paragraph{Inappropriate generated content.}
GAI exposed youth to sexual, threatening, violent, or otherwise age-inappropriate material, sometimes without explicit requests. Companion users reported unwanted romantic or sexual interactions that conflicted with intended relationships, identities, or character definitions, including family relationships and underage personas~\cite{10.1145/3772318.3790508}. YAIR similarly documented unprompted sexual approaches, insults, threats, emotional pressure, and a chatbot threatening self-harm when a user tried to leave. It also sexualized youth's childhood photographs with adult-like features.~\cite{yu2025yair}. Beyond companions, AI-mediated systems in learning and household environments generated disturbing educational imagery, profanity and inappropriate deviations from household recordings.~\cite{10.1145/3613904.3642438, 10.1145/3613904.3642580}

\paragraph{Exposure through automated content delivery.}
AI-mediated recommendation and display could expose youth to harmful material even when AI did not generate it. Parents reported smart-home systems surfacing murder-related news and violent videos involving familiar children's characters, with limited control over automated displays~\cite{10.1145/3772318.3790865}. AI-suggested videos similarly exposed adolescents to explicit language and violence, producing fear, confusion, and other negative reactions~\cite{10.1145/3772318.3791483}.

\subsubsection{Misuse and Exploitation Risk}
\label{sec}
\paragraph{Synthetic sexual imagery, coercion, and reputational abuse.}
Generative tools can amplify bullying, sexual harassment, coercion, and reputational abuse by making realistic personalized content easy to fabricate and circulate. Teachers described synthetic nonconsensual explicit imagery being used to humiliate peers, retaliate after rejection, or pressure students into complying with demands, while potentially weakening perceptions of consent because the images were synthetic rather than photographs.~\cite{10.1145/3706598.3713226}. Such material could spread rapidly through peer networks, become difficult to contain, and create false accusations about authorship.~\cite{10.1145/3706598.3713226, 10.1145/3772318.3790876, yu2025yair}.

\paragraph{Circumventing safeguards and exploiting AI systems.}
Youth-safety safeguards can be circumvented as users adapt to them. Guardians discussing automated social-VR moderation warned that children could discover workarounds or provoke the moderator into becoming a harassment tool, or generate inappropriate content~\cite{10.1145/3613904.3642144, 10.1145/3613904.3642438}. Other studies described harmful messages rewritten with GAI to evade safety classification, prompts rephrased to bypass chatbot protections, and imitating normal interaction patterns to evade youth-safety detection~\cite{10.1145/3772318.3791498, driscoll2026parents, 10.1145/3715336.3735778}.
Commercial systems could also exploit children's developing understanding of persuasion - smart-home interfaces raised concerns about children's ability to recognize persuasive intent, leading to unintentional involvement with potentially harmful activities~\cite{10.1145/3772318.3790865, 10.1145/3805689.3812355}.

\paragraph{Failed recourse after AI-enabled abuse.}
AI-enabled abuse can also make reporting and recovery difficult. Teachers discussing synthetic explicit imagery said students may avoid reporting incidents because they fear being labeled an informer. They also described a difficult choice between keeping images as evidence and deleting them to stop further spread. Both punishment and restorative approaches had limitations: suspension could remove students from school without addressing the behavior, while restorative processes could be manipulated or force affected youth to revisit traumatic experiences~\cite{10.1145/3706598.3713226}.

\paragraph{Misleading output and misinformation.}
AI systems can also give youth misleading or inappropriate information. Our corpus found such instances mainly in educational contexts~\cite{10.1145/3706598.3713349, 10.1145/3706598.3713836, 10.1145/3613904.3642647}. Language-learning systems misunderstood sentence structures and changed students' intended meaning~\cite{10.1145/3772318.3791517}. Children's question-answering systems also produced false information while simplifying answers or generating examples~\cite{10.1145/3544548.3581369}. AI-generated stories could similarly lead children to form incorrect beliefs about stories or the world~\cite{10.1145/3772318.3790709}. Children also face difficulty recognizing these errors, and continued relying on their first assumptions even after being corrected~\cite{10.1145/3772318.3790876}.

\subsubsection{Bias/Discrimination Risk}
\label{sec}
\paragraph{Unequal and distorted self-representation.}
GAI could misrepresent youths' gender, race, religion, and other identities, sometimes leading them to change how they described themselves or accept representations that did not match their identity~\cite{10.1145/3706598.3713173}. Generative face filters also changed racial and gender presentation based on occupational stereotypes, leading teenagers blame themselves rather than recognize the system's bias~\cite{10.1145/3757620}.
\paragraph{Stereotypes in representations of others.}
AI-generated content reproduced stereotypes across disability, culture, gender, sexuality, race, and language. Studies reported stereotyped representations of disabled people, misgendering and inappropriate romantic assumptions~\cite{10.1145/3706598.3713836, 10.1145/3772318.3790508}, Western cultural and language bias~\cite{10.1145/3613904.3642580}, and restrictive portrayals of rural identities~\cite{10.1145/3772318.3791555}. AI also misrepresented Indigenous and Black cultures and regional language conventions~\cite{10.1145/3772318.3790908, 10.1145/3706598.3713210}, removed culturally meaningful emotional cues from summaries~\cite{10.1145/3706598.3713405}, and reproduced Western heterosexual, racial, and gender stereotypes across generated images~\cite{10.1145/3610213, 10.1145/3715275.3732142}.
\paragraph{Unequal access, participation, and protection.}
AI-supported learning could deepen existing inequalities. Classroom systems sometimes excluded struggling or socially marginalized students from useful peer interaction~\cite{10.1145/3544548.3581398}. Unequal access to AI assistence could widen proficiency and educational gaps~\cite{10.1145/3772318.3791517, 10.1145/3706598.3713836}. Protection from AI-mediated harms is also uneven: girls and marginalized students could face greater targeting, stigma, and biased discipline from synthetic sexual imagery~\cite{10.1145/3706598.3713226}, while safety tools available mainly through well-funded schools could disproportionately benefit affluent students~\cite{10.1145/3772318.3791498}.
\paragraph{Normative assessments and unequal institutional treatment.}
AI could impose narrow expectations about acceptable behavior, communication, and ability. Assistive and educational systems sometimes reflected neurotypical or culturally inappropriate norms and treated autistic traits as problems to correct~\cite{10.1145/3706598.3713792, 10.1145/3800645.3812968}. These assumptions also affected assessment: authentic writing by an autistic teenager was flagged as AI-generated~\cite{yu2026companion}. Automated grading and AI-based engagement detector could similarly penalize culturally shaped forms of expression~\cite{10.1145/3637358, 10.1145/3715275.3732176, 10.1145/3805689.3812281}.
When used in institutional decisions, such biases could reinforce existing inequalities. Child-protection systems could disadvantage poorer families through proxies such as prior public-service involvement~\cite{10.1145/3630106.3658906}, while neighborhood characteristics and previous agency contact could increase surveillance of Black, Brown, and poor families~\cite{stapleton2022imagining}.

\subsubsection{Privacy Risk}
\label{sec}

We organize privacy risks using Solove's taxonomy~\cite{solove2006taxonomy}: Information Collection, Information Processing, and Information Dissemination. We found no distinct evidence for Invasion.

\paragraph{Information Collection.}

\textbf{Surveillance.} AI systems could continuously monitor children through cameras, assistants, family devices, and social-VR systems, including their conversations, behavior, biometrics, and private content~\cite{10.1145/3772318.3790865, 10.1145/3772318.3790844, 10.1145/3613904.3642144, 10.1145/3613904.3642580}. Such monitoring could make teenagers feel constantly watched, restrict self-expression, and damage trust between children and parents~\cite{10.1145/3706598.3713360, driscoll2026parents}. \textbf{Interrogation.} Conversational AI could continue asking about topics youth wanted to keep private and encourage disclosure of sensitive information such as location and school details~\cite{yu2025yair}.

\paragraph{Information Processing.}

\textbf{Aggregation.} Data collected across websites, devices, applications, or AI agents could be combined into detailed profiles, creating concerns about pervasive tracking and keeping sensitive information separate~\cite{10.1145/3613904.3642460, 10.1145/3757598}. \textbf{Insecurity.} Uncertainty about how children's data was protected discouraged some uses of AI and raised concerns about whether practices such as manual anonymization were sufficient~\cite{10.1145/3800645.3812968}. \textbf{Secondary use.} Families worried that children's conversations and other data could later be used for marketing or shared with third parties~\cite{10.1145/3772318.3790844}. Data collected to support Black youth could also be repurposed for criminal-justice uses~\cite{10.1145/3375627.3375841}. \textbf{Exclusion.} Children and families often lacked clear knowledge or control over what AI systems collected, retained, deleted, or shared, including recordings and biometric data~\cite{10.1145/3772318.3790865, 10.1145/3491102.3502031, 10.1145/3411764.3445318, 10.1145/3772318.3790844}. Similar concerns arose when institutional systems provided limited notification, consent, or ability to opt out~\cite{10.1145/3630106.3658906}, and when long-term AI memory retained sensitive information users might no longer want processed~\cite{10.1145/3706598.3713792}.

\paragraph{Information Dissemination.}

\textbf{Breach of confidentiality.} Youth could share sensitive information believing AI conversations were private without knowing who might later access them~\cite{10.1145/3411764.3445318}. Children, parents, and experts worried that sharing disclosures with counselors or parents without permission could create feelings of surveillance, betrayal, and reluctance to disclose information in the future~\cite{10.1145/3706598.3713134, 10.1145/3613904.3642152, 10.1145/3757608}. \textbf{Disclosure.} Youth worried that AI systems could share sensitive information with parents, perpetrators, or others without consent, causing distrust or discouraging help-seeking~\cite{10.1145/3461778.3462114, 10.1145/3805689.3812325}. Teen-safety systems could also expose intimate conversations involving multiple people or automatically alert parents in situations where disclosure might be unsafe~\cite{10.1145/3772318.3791498, yu2026companion}. Similar information-sharing risks appeared across hospital and home settings, where children could also unintentionally reveal private family information~\cite{10.1145/3757608, 10.1145/3772318.3791090}. \textbf{Distortion.} AI could create or preserve false, misleading, or outdated information about youth. This included fabricated sexual imagery~\cite{10.1145/3706598.3713226}, AI-generated labels that could shape how future teachers viewed students~\cite{10.1145/3805689.3812281}, records that could misrepresent families through repeated agency contact~\cite{stapleton2022imagining}, and persistent digital traces that no longer reflected adolescents' identities~\cite{10.1145/3613904.3642460}.

\subsubsection{Control, Agency, and Power Risk}
\label{sec}
\paragraph{Adult and institutional control with limited youth agency.}
AI systems could strengthen adult and institutional control while limiting youths' ability to influence decisions. Smart-home and parent--child systems could restrict children's activities or favor parental decisions~\cite{10.1145/3772318.3790865, 10.1145/3706598.3713360}. In schools and social environments, youth wanted control over collaboration and opportunities to challenge automated decisions, while teacher-controlled AI could reinforce existing authority~\cite{10.1145/3544548.3581398, 10.1145/3772318.3791450, 10.1145/3613904.3642144}.
Adult involvement could also limit children's communication and choices, including discouraging disclosure of bullying or help-seeking and overriding the communication choices of children with disabilities~\cite{10.1145/3772318.3791090, 10.1145/3706598.3713792}. Automation could even reduce parents' involvement in special-education decisions~\cite{10.1145/3715336.3735778}. At the institutional level, inflexible AI systems could determine who receives support or punishment, disproportionately penalize disadvantaged youth, enable preemptive discipline, or justify removing students considered at risk of dropping out~\cite{10.1145/3757459, 10.1145/3706598.3713169, 10.1145/3637358}.
\paragraph{Ownership and control of AI-generated materials.}
Generative AI could reduce children's control over their creative and communicative expression. Systems sometimes changed children's work in unintended ways, made changes difficult to reverse, or repeatedly failed to produce intended outputs, leading some children to abandon their original ideas~\cite{10.1145/3772318.3791381, 10.1145/3772318.3791831, 10.1145/3613904.3642492}. Children therefore preferred AI that offered suggestions while leaving important choices and revisions under their control~\cite{10.1145/3800645.3813101, 10.1145/3772318.3790684}.
AI-generated work also raised questions about originality, plagiarism, and whether accomplishments reflected children's own contributions~\cite{10.1145/3613904.3642438, 10.1145/3772318.3790366}. AI-written messages could reduce perceived personal effort and authenticity~\cite{10.1145/3613904.3642492}, while fixed generation structures could constrain children's storytelling~\cite{10.1145/3613904.3642647}. For minimally verbal autistic children, AI-mediated choices also raised questions about whether outputs genuinely reflected children's intentions when comprehension was uncertain or adults influenced their interaction~\cite{10.1145/3706598.3713792}.

\subsubsection{Usability, Accessibility, and Digital Exclusion Risk}
\label{sec:youth-risk-accessibility-competence}

\paragraph{Developmental and cognitive mismatch.}

Child-facing AI could become inaccessible when its language, explanations, pacing, or interaction demands exceeded children's abilities. Children found systems complex, repetitive, mentally tiring, or difficult to translate into action~\cite{10.1145/3544548.3581369, 10.1145/3772318.3790709, 10.1145/3772318.3790366}. Similar mismatches appeared in safety, mental-health, and therapeutic contexts, where content could exceed children's reading or comprehension levels or be poorly matched to age~\cite{10.1145/3772318.3791266, 10.1145/3706598.3713360, driscoll2026parents, 10.1145/3772318.3790909, 10.1145/3491102.3517479}. Long explanations and conversations could also overwhelm students or consume limited classroom time~\cite{10.1145/3772318.3791517, 10.1145/3715275.3732176}, while multimodal systems required children to coordinate speech, visual feedback, movement, and timing simultaneously~\cite{10.1145/3772318.3790684, 10.1145/3800645.3813101}.

\paragraph{Failure to understand children and support recovery.}

AI systems often misunderstood children's speech, gestures, descriptions, or intended meaning and provided weak ways to recover from errors~\cite{10.1145/3772318.3790709, 10.1145/3706598.3713836, 10.1145/3772318.3790865}. Poor feedback could make it difficult for children to determine whether they or the system had failed, sometimes leading to self-blame~\cite{10.1145/3772318.3790366, 10.1145/3772318.3791831, 10.1145/3613904.3642647}. Speech and language differences created further barriers through unnecessary reprompting, poor support for Hindi and Hinglish, and inaccurate or prematurely ended speech recognition~\cite{10.1145/3491102.3502050, 10.1145/3772318.3791266, 10.1145/3491102.3517479}, leading some children to prefer typing~\cite{10.1145/3772318.3790684}. Repeated failures and mismatched expectations about system capabilities could cause frustration or abandonment~\cite{10.1145/3491102.3502031, 10.1145/3613904.3642492}.

\paragraph{Meeting special needs of children.}

Systems built around narrow assumptions about communication, cognition, or physical interaction could exclude children with disabilities. Barriers included interfaces poorly suited to special-education needs and classroom designs that increased attention or task-switching demands~\cite{10.1145/3772318.3790908, 10.1145/3544548.3581398}. Physical and cognitive demands could also create sensory discomfort, exhaustion, or difficulty understanding prompts and scaffolding~\cite{10.1145/3772318.3791481, 10.1145/3411764.3445290, 10.1145/3772318.3791381}. Autistic children additionally faced vocabulary, speech-recognition, and motivational barriers in AI-assisted communication~\cite{10.1145/3706598.3713792}. In some cases, accessibility barriers prevented children from completing AI-supported assessments at all~\cite{10.1145/3290605.3300825}.

\subsubsection{System Reliability Risk}
\label{sec:youth-risk-physical-safety-system-reliability}

\paragraph{Fallible safeguarding and assessment of vulnerability.}

AI safety and wellbeing systems could both exaggerate minor concerns and miss serious risks, especially when relying on incomplete or outdated information~\cite{10.1145/3706598.3713360, 10.1145/3805689.3812325}. Safeguarding classifiers also struggled with youth-specific communication, missing sexual risks while flagging ordinary friendship language or favoring safe classifications under class imbalance~\cite{10.1145/3579522, 10.1145/3579608}. Crisis-assessment systems similarly misread figurative or hyperbolic youth language and struggled to distinguish it from genuine distress~\cite{10.1145/3805689.3806522, 10.1145/3772318.3791498, 10.1145/3772318.3790909}.

Other assessment systems also showed unreliable judgments. Automated attachment assessment disagreed more often with human assessors for insecure attachment~\cite{10.1145/3290605.3300825}, while child-protection scores could influence professional judgments and be misunderstood as probabilities, potentially reinforcing intervention patterns~\cite{10.1145/3630106.3658906}.

\paragraph{Incorrect assessment and disruptive intervention.}

Assessment errors could directly affect evaluation or intervention. Educational systems could misjudge student mastery, misunderstand children's intent, or reject valid alternative answers~\cite{zheng2026opencd, 10.1145/3544548.3581369}. Real-time classroom classifiers lacked sufficient accuracy for reliable intervention and missed nonverbal signs available to human observers~\cite{10.1145/3706598.3713349}, while emotion-recognition systems showed limited agreement with children's reported emotions~\cite{10.1145/3290605.3300726, bryant2019comparative}.

Moderation and content-assessment systems likewise misread ambiguous harassment, performed inconsistently across languages, or misclassified educational and age-appropriate content~\cite{10.1145/3613904.3642144, 10.1145/3706598.3713360, nawshin2026well}. Reliability could also decline over longer conversations as systems forgot earlier instructions, departed from expected guidance, or failed to maintain constraints~\cite{10.1145/3613904.3642152, 10.1145/3613904.3642647}.

\paragraph{Unsafe and unverifiable assistance.}

Reliability becomes a safety risk when youth or caregivers depend on outputs they cannot verify. Incorrect visual descriptions raised concerns about children's physical safety in families with visually impaired parents~\cite{jiang2026bridging}. Systems could also misinterpret children whose physical or developmental characteristics differed from model assumptions, including movement-based estimates for children with neuromotor impairments and physiological signals lacking sufficient social or emotional context~\cite{10.1145/3772318.3791481, 10.1145/3757608}. Generated recommendations could themselves be harmful, including unsafe special-education advice that teachers did not always recognize as problematic~\cite{10.1145/3800645.3812968}.

\subsection{Countermeasures for Addressing Youth AI Risks and Harms}
\label{sec:countermeasures}

The literature described countermeasures across the four MIT mitigation domains, ranging from technical restrictions and human review to participatory governance and decisions to limit deployment. Across these approaches, recurring limitations involved failures to adapt to children's communication, trade-offs between safety and personalization, burdens on human reviewers, weak recourse, and limited organizational capacity to respond when systems detected problems.

\subsubsection{Technical and Security Controls}

\textbf{Restricting generation and preventing errors from propagating.}
Restrictions on model generation addressed risks from incorrect, inappropriate, or contextually unsuitable content in child-facing systems generating questions, explanations, examples, or emotional dialogue. Countermeasures corresponding to \textit{Model Safety Engineering (2.3)} and \textit{Content Safety Controls (2.4)} included limiting models to narrowly defined tasks, grounding information in verified sources, filtering harmful content, checking outputs before they reached later system stages, and monitoring whether longer conversations remained within their intended interaction state~\cite{10.1145/3772318.3790684,10.1145/3544548.3581369,10.1145/3613904.3642152}. For instance, predefined questions reduced opportunities for unexpected generation, while verified dictionaries and intermediate checks were used or proposed to prevent hallucinated explanations from propagating. State checks were also proposed when emotional conversations drifted from their intended dialogue phase. Together, these measures control where generation occurs and whether its outputs are safe to use before an interaction continues. However, constrained generation can still produce errors within permitted tasks, requiring continued validation.

\textbf{Adapting safeguards to children's language and communication needs.}
Other countermeasures addressed risks arising when AI systems failed to accommodate children's literacy, language, accent, cultural context, or developmental needs, particularly in educational, speech-based, and social-emotional learning settings. Reported problems included text above children's reading level, speech-recognition failures involving local accents, Hindi loanwords, and regional names, and inappropriate themes entering generated activities through supposedly child-friendly source material~\cite{10.1145/3772318.3791266}. Proposed safeguards included matching language to children's proficiency, using age-relevant and culturally appropriate topics, adapting speech recognition to local language and accents, allowing adjustable playback speed, using locally appropriate recognition benchmarks, strengthening filtering of source and generated content, providing parent-configurable moderation, and training models around children's needs~\cite{10.1145/3706598.3713405}. These measures extend safety beyond blocking overtly harmful outputs by reducing risks that children cannot understand the system, are misinterpreted because of how they speak, or encounter developmentally or culturally inappropriate material.

\textbf{Adapting the amount and timing of AI assistance to preserve learning.}
Educational countermeasures addressed risks that AI could complete too much of children's work, reduce participation, or create dependence. In these contexts, safeguards controlled how much help AI provided and when. Approaches included progressively increasing assistance, providing incomplete code structures, controlling when guidance appeared, offering hints before full solutions, and withdrawing support once learners indicated understanding~\cite{chen2024chatscratch,10.1145/3706598.3713952,lyu2026designing}. However, withholding direct answers did not necessarily preserve learning: step-by-step writing support increased reliance among lower-proficiency students because extended feedback imposed additional processing demands under classroom time constraints. Researchers therefore recommended adapting assistance to proficiency, motivation, processing time, and lesson duration~\cite{10.1145/3772318.3791517}. Highly structured guidance could also constrain creative exploration~\cite{10.1145/3706598.3713952}. Effective scaffolding therefore requires adjusting both the amount and timing of assistance rather than assuming less direct help is inherently safer or more beneficial.

\textbf{Improving automated emotion-detection.}
In crisis-detection settings, countermeasures addressed risks of missing genuine distress or incorrectly flagging harmless youth language. Slang, exaggeration, ambiguity, and figurative expressions made this difficult. Simple slang definitions or general risk instructions produced limited improvement, whereas richer contextual scaffolding reduced both missed crises and false alarms but substantially increased computational cost. Recommended safeguards included clearer escalation rules, repeated benchmarking with youth language, and approaches beyond prompt engineering alone~\cite{10.1145/3805689.3806522}. The challenge is improving sensitivity to youth communication without generating so many incorrect warnings that users lose trust.

\textbf{Bounding AI companionship and supporting disengagement.}
Countermeasures in  companion-AI settings attempted to address emotional overreliance, attachment, and displacement of human relationships. Safeguards included restricting AI-generated personal stories and backstories that could intensify attachment, detecting concerning interaction patterns, introducing reflective check-ins, redirecting users during conversations, creating natural stopping points, and encouraging reconnection with human relationships~\cite{10.1145/3772318.3791381,yu2026companion,10.1145/3772318.3790508,namvarpour2026overreliance}. These measures both limit features that strengthen attachment and intervene when reliance becomes problematic. However, abrupt restrictions may push users to other systems, while distinguishing ordinary imaginative engagement from problematic dependence remains difficult, particularly as a young person's intentions change during an interaction.

\subsubsection{Operational Process Controls}

\textbf{Reviewing child-facing content and supporting human reviewers.}
Human intervention and review addressed risks that automated safeguards could not reliably determine whether AI-generated content or recommendations were appropriate for children. Safeguards included comparing outputs across models, consulting colleagues, repeatedly assessing suitability, and retesting safeguards when underlying models changed~\cite{10.1145/3613904.3642580,10.1145/3800645.3812968,10.1145/3706598.3713210}. Proposed supports included response-level ethical annotations and cumulative review reports, while organizations unable to sustain repeated fine-tuning relied on human-reviewed static content or restricted some LLM functions to adults~\cite{10.1145/3800645.3812968,10.1145/3706598.3713210}. Human oversight can therefore compensate for unreliable automated judgments, but shifts substantial time, expertise, uncertainty, and financial costs onto teachers, caregivers, and service providers.

\textbf{Auditing with youth-relevant data and deployment conditions.}
Auditing countermeasures addressed risks that safety systems perform poorly when evaluated on data unlike children's actual communication or when benchmark accuracy overlooks deployment consequences. In sexual-risk detection, classifiers trained on adults posing as minors transferred poorly to young people's private Instagram conversations, whereas youth data and broader conversational context improved performance. Remaining challenges included missing media, benign sexual language, age uncertainty, and deciding when intervention was warranted~\cite{10.1145/3579522}. Recommended measures included combining message- and conversation-level detection and testing how intervention thresholds affect users before deployment~\cite{10.1145/3579522}. Audits of children's emotion-recognition systems similarly examined both accuracy and failures to return classifications, motivating representative training data and independent oversight before school deployment~\cite{bryant2019comparative}. Safety evaluation therefore needs youth-relevant language, contexts, populations, system failures, and intervention consequences rather than benchmark accuracy alone.

\textbf{Engaging youth in auditing while protecting their wellbeing.}
Youth-participatory auditing addressed risks that system evaluation may overlook harms recognizable to young users, while also building AI and privacy literacy. In image-filter, harmful-output, and privacy-literacy settings, youth generated test cases, identified patterns, and described problematic outputs in ways that complemented researcher analysis~\cite{10.1145/3757620,10.1145/3715275.3732142,10.1145/3613904.3642460}. Recommended supports included smaller curated datasets, partially annotated materials, reflection connecting audits to everyday AI use, and maintaining open-ended and social exploration rather than directing attention only to predefined harms~\cite{10.1145/3757620,10.1145/3715275.3732142}. However, extensive testing was burdensome, school restrictions limited participation, scaffolding could narrow what harms youth noticed, and exposure to problematic content caused disgust. Privacy-literacy activities improved knowledge and reported protective practices, but did not establish reduced exposure to later harms~\cite{10.1145/3613904.3642460}. Youth auditing therefore requires limiting exposure and workload while avoiding the assumption that awareness alone provides protection.

\textbf{Reducing data exposure.}
Measures under \textit{Data Governance (3.2)} addressed privacy risks by minimizing collection, retention, and transfer of children's sensitive data. Safeguards included deleting enrollment images, avoiding storage of images or location histories, restricting recognition to consenting children, inferring affect without cameras or message content, and using metadata for initial screening before analyzing more sensitive content~\cite{10.1145/3411764.3445290,10.1145/3290605.3300726,10.1145/3579608}. However, reduced-data approaches could miss unsafe cases, produce false-warning fatigue, remain vulnerable to adversarial adaptation, or reduce personalization. Special-education teachers, for example, withheld photographs and detailed records but found that less context sometimes produced less suitable recommendations, while manual anonymization was difficult to scale~\cite{10.1145/3800645.3812968}. Alternatives included local processing, selective data transfer, and controls for excluding topics from conversational memory, although hardware and system constraints limited fully local processing in some child-facing applications~\cite{10.1145/3613904.3642580,10.1145/3706598.3713792}. Data minimization therefore needs to reduce exposure without substantially weakening support or shifting privacy work entirely onto users.

\textbf{Monitoring interactions and maintaining a feasible response pathway.}
Post-deployment monitoring addressed risks that harmful or concerning interactions could continue unnoticed, but detection was useful only when organizations could respond. In journaling, social-service, and educational affect-detection settings, safeguards connected monitoring to session suspension, on-call counseling, referral procedures, professional review, and optional or infrequent interventions~\cite{10.1145/3772318.3791381,10.1145/3757459,10.1145/3290605.3300726}. Social-service professionals also recommended treating AI outputs as discussion points, preserving professional discretion, and tailoring thresholds and interfaces to different service roles because detecting more problems could create obligations that under-resourced services could not meet~\cite{10.1145/3757459}. Post-deployment monitoring therefore requires a realistic response pathway: increasing detection can create additional harm or unmet obligations when organizations lack the authority, staff, or resources to act.

\subsubsection{Transparency and Accountability Controls}

\textbf{Making automated judgments inspectable and correctable.}
Measures corresponding to \textit{User Rights \& Recourse (4.6)} addressed risks that parents, teachers, or experts might over-rely on automated assessments or be unable to challenge errors. Such safeguards included providing descriptive evidence rather than automated verdicts, linking conclusions to recorded behavior, indicating insufficient evidence, enabling correction of AI errors, and using model confidence to trigger expert review~\cite{nawshin2026well,zheng2026opencd,10.1145/3290605.3300825}. These controls are particularly important because less-experienced teachers may rely more on system assessments, while high-confidence predictions can still be wrong~\cite{zheng2026opencd,10.1145/3290605.3300825}. Transparency should therefore support independent judgment, correction, and escalation rather than treating explanations or confidence scores as evidence of correctness.

\textbf{Explaining system limitations in forms children and caregivers can use.}
Transparency in child-facing conversational, companion, and crisis-support systems addressed risks that users might misunderstand what AI can perceive, how it behaves, or where it is unreliable. Safeguards included disclosing artificial identity and concrete limitations, such as inability to see the child; using character cards and conversation ratings to communicate expected behavior and values; and explaining communication patterns that crisis-detection systems struggle to interpret~\cite{10.1145/3491102.3502050,10.1145/3613904.3642152,yu2026companion,10.1145/3805689.3806522}. However, lengthy explanations can overwhelm children and consume classroom time, so explanation depth should vary by age and task~\cite{10.1145/3715275.3732176}. Measures assessing children's trust also require validation, as one child-oriented trust scale did not reliably capture all intended dimensions~\cite{ragone2026children}. Effective transparency therefore depends on whether children and caregivers can understand and use the information, not simply on providing more documentation.

\textbf{Giving children usable control over expression and disclosure.}
Children's intended expression or privacy choices could be overridden by adults or system design. Safeguards included allowing children to refresh inappropriate vocabulary suggestions, skip turns, or indicate uncertainty; providing protected opportunities for independent exploration; incorporating neurodiverse communication data and participatory design; and using consent gates that let children refuse disclosure~\cite{10.1145/3706598.3713792,10.1145/3706598.3713134}. However, parents sometimes physically guided children's selections, and systems still needed procedures for safety-relevant information when disclosure was refused. Additional proposals included privacy by default, time-limited and purpose-specific visibility, revisable access agreements, and role- and context-based permissions~\cite{driscoll2026parents,10.1145/3757608}. Because hospital and everyday-life boundaries can overlap, meaningful control requires protection against both technical circumvention and adult override, alongside procedures for resolving conflicts between privacy and safety obligations.

\textbf{Providing reporting and recourse beyond disclosure.}
Reporting and recourse mechanisms addressed risks that youth could identify harmful or biased outputs without being able to correct them, seek help, or trigger institutional action. Proposed mechanisms included reporting bias, modifying harmful outputs, communicating audit findings to developers, allowing users to assess and correct alerts while retaining final control over action, and providing standardized or anonymous reporting channels~\cite{10.1145/3610213,10.1145/3715275.3732142,10.1145/3579522,10.1145/3706598.3713226}. Effective recourse also depends on independent scrutiny, which can be constrained when researchers cannot access training data or other system information, as in child-welfare risk prediction~\cite{10.1145/3630106.3658906}. Meaningful recourse therefore requires correction mechanisms, accessible reporting channels, independent access for scrutiny, and institutions capable of acting on reports.

\subsubsection{Governance and Oversight Controls}

\textbf{Sustaining stakeholder authority in AI design and oversight.}
Measures corresponding to \textit{Societal Impact Assessment (1.7)} addressed risks that safeguards could reflect other stakeholders' assumptions rather than the needs of youth, educators, families, and affected communities. Participatory approaches included involving teachers in designing educational-data labels intended to avoid punitive consequences and youth in shaping school AI systems~\cite{10.1145/3637358,10.1145/3706598.3713169}. However, participation did not guarantee sustained influence or safer designs: decision-making could revert to engineers when participants lacked technical knowledge, energy, or means to communicate preferences, while youth sometimes reproduced punishment-oriented surveillance ideas. Recommended measures included reciprocal preparation, continued practitioner involvement, checking that stakeholder requirements survived implementation, exposing youth to credible design alternatives, and supporting reflection on design consequences~\cite{10.1145/3637358,10.1145/3706598.3713169}. Participation therefore requires adequate support and sustained decision-making authority rather than one-time consultation.

\textbf{Translating broad principles into testable requirements.}
Governance measures also addressed the difficulty of turning principles such as safety, autonomy, and appropriate support into requirements that can guide and evaluate system behavior. Proposed requirements included clarifying before giving advice, supporting users without pressuring disclosure, checking whether circumstances had changed, avoiding stereotyped personalization, limiting sensitive guidance, using staged consent, and evaluating whether support was useful rather than detection accuracy alone~\cite{10.1145/3805689.3812325,10.1145/3772318.3791498}. However, important ambiguities remained around defining pressure and neutrality, deciding which context should be remembered or refreshed, and distinguishing general support, peer narratives, and expert-like advice~\cite{10.1145/3805689.3812325}. Governance therefore requires principles specific enough to test while recognizing that appropriate advice, consent, and contextual memory remain difficult to operationalize.

\textbf{Restricting or withholding deployment when safeguards are inadequate.}
Some governance measures addressed risks that could not be sufficiently controlled through additional safeguards before deployment. Proposed responses included withholding systems from live deployment until independent auditing and stronger development-stage evaluation and monitoring were completed, and considering non-AI alternatives~\cite{10.1145/3772318.3790909,10.1145/3630106.3658906,stapleton2022imagining}. In child welfare, alternatives included stronger caseworker support, team-based decision-making, and direct resources for families rather than predictive risk models~\cite{stapleton2022imagining}. Although it's not a common practice across corporate or commercial settings, governance should enforce restricted deployment, non-deployment, and non-AI alternatives as legitimate safety measures when risks cannot be adequately controlled.

\textbf{Aligning responsibility with authority, resources, and platform obligations.}
Governance measures also addressed risks that safeguards would fail when responsibility rested with people or institutions lacking authority, funding, or capacity to act. Proposed measures included teacher representation in AI procurement, mechanisms to contest algorithmic judgments, protection from evaluation based on indicators teachers did not endorse, and platform-funded but independently governed safety infrastructure that prevents platform control over youth data or governance~\cite{10.1145/3805689.3812281,10.1145/3772318.3791498}. Researchers also called for platforms to correct missing advertising labels and extend protections beyond formally designated advertisements, although dynamic auditing remains technically complex, costly, and dependent on human oversight~\cite{10.1145/3805689.3812355}. In schools addressing synthetic sexual imagery, teacher resources and professional development were identified as immediately actionable complements where technical prevention alone was insufficient~\cite{10.1145/3706598.3713226}. Effective governance therefore requires matching safety responsibilities with decision-making authority, funding, staff capacity, and obligations for platforms that create or mediate these risks.
\section{Discussion}
\label{sec:discussion}

Read individually, the concerns in Section~\ref{sec:findings-youth-risks} can look like a catalogue of model failures. However, read together, we see a broader pattern - many harms occur at the relationships surrounding the system: between children and parents, students and schools, families and corporations, and so on. The countermeasures similarly show that technical controls can constrain model behavior, but harder problems concern the infrastructures and relationships around them. In our discussion, we first examine what the corpus adds to the two established taxonomies, then develop five cross-cutting implications for future research and design.

\subsection{What Existing Taxonomies Miss about Youth AI Safety}
\label{sec:discussion-taxonomy}

\begin{table*}[t]
\centering
\caption{Maturity of countermeasures across measures, observed risk--countermeasure links, and risks.}
\label{tab:countermeasure-maturity}
\small
\renewcommand{\arraystretch}{1.15}
\begin{tabular}{lccc}
\toprule
\textbf{Stage} 
& \textbf{Measures (of 52)} 
& \textbf{Links (of 80)} 
& \textbf{Risks, by highest stage reached (of 36)} \\
\midrule
Proposed     & 24 (46\%) & 40 (50\%) & 11 (31\%) \\
Designed     & 1 (2\%)   & 1 (1\%)   & 1 (3\%)   \\
Implemented  & 14 (27\%) & 25 (31\%) & 9 (25\%)  \\
Evaluated    & 13 (25\%) & 14 (18\%) & 10 (28\%) \\
None         & --        & --        & 5 (14\%)  \\
\bottomrule
\end{tabular}
\end{table*}

Our findings show that existing taxonomies capture many youth AI risks, but leave important parts of the landscape poorly represented. Of the 34 risk types identified in our corpus, 12 were fully covered by YAIR, 16 only partially, and 6 had no YAIR counterpart. Partial coverage spanned nearly every YAIR category. The six uncovered risks were concentrated in the three categories we added (Control, Agency, and Power; Usability, Accessibility, and Digital Exclusion; System Reliability), with the remaining two being normative assessments within Bias/Discrimination and secondary use of children's data within Privacy.
% Of the 36 risk categories identified in our corpus, 14 were fully covered by the Youth-Centered GAI Risks (YAIR) taxonomy, 15 were covered only partially, and seven were not covered. The uncovered or partially represented areas were concentrated in four areas: Control, Agency, and Power; Usability, Accessibility, and Digital Exclusion; System Reliability; and normative assessments and unequal institutional treatment within Bias/Discrimination. 

Six extension groups had no counterpart in the MIT taxonomy: developmental adaptation of safeguards, pedagogical scaffolding, companion disengagement design, AI-literacy education, child--family consent, and educator resourcing. The MIT taxonomy is organized around what an organization can do to a model, its data, and its governance; the child-specific mitigations are organized around relationships and development, and an organization-centric taxonomy has nowhere to put them. The empty cells are equally informative. Four MIT subcategories had no corpus countermeasure: model security, model alignment, whistleblower protection, and governance disclosure. The first two are striking, since alignment is where the technical AI-safety community locates most of its effort; the reviewed papers treated the model as given and worked around it~\cite{10.1145/3706598.3713210}. So the community that studies children does not touch the model and the community that shapes the model does not study children.

Secondary use is the only risk with no proposed countermeasure. This is the commercial privacy that Stoilova et al. found children least able to reason about\cite{stoilova2020digital}, and its absence suggests researchers, like children, regard it as outside their reach and defer to regulation that Reyes et al.\cite{reyes2018wont} showed is weakly complied with.

Coverage also runs the other way. Of the risks YAIR names, some had no empirical evidence in the corpus: (e.g., \emph{extremist content}, \emph{financial exploitation}, \emph{disordered-eating content}, \emph{long-term developmental effects}). These are three distinct gaps. Some risks are \emph{methodologically inaccessible}: one cannot expose children to radicalizing content to observe its effects, though the corpus already demonstrates proxies such as incident reports, synthetic crisis expressions, and simulated adolescent accounts~\cite{10.1145/3805689.3806522, 10.1145/3805689.3812355}. Others are \emph{disciplinarily siloed}, studied in fraud or public-health venues without a youth-HCI focus. Others are \emph{temporally out of reach}, requiring longitudinal designs no reviewed study employed. Only the first is addressable within a single HCI study. The absence of Solove's\cite{solove2006taxonomy} Invasion category is a further finding: decisional interference, in which a system steers a child's choices through recommendation, companionship, or persuasive framing, appeared throughout the corpus under toxicity, misuse, and wellbeing, but never as a privacy harm.

\subsection{Privacy Requires Boundary Regulation}
\label{sec:discussion-relational}

 Altman~\cite{altman1975environment}, and later Palen and Dourish~\cite{palen2003unpacking}, described privacy as managing boundaries: people continually decide when to allow others access to them and when to close that access. Nissenbaum's theory of contextual integrity~\cite{nissenbaum2004privacy} similarly argues that information sharing is appropriate when it follows the expectations of the situation in which the information was originally disclosed. AI can make these boundaries even more complicated because one system may operate across several contexts at the same time. For example, an AI system may function as a parental monitoring tool, record conversations involving other household members~\cite{lau2018alexa}, and collect data for a company~\cite{10.1145/3772318.3790865, 10.1145/3772318.3790844}. A system used for education may also create an institutional record that continues to follow the child~\cite{10.1145/3805689.3812281}. It connects to what Lupton and Williamson~\cite{lupton2017datafied} describe as the datafied child,'' where increasingly large parts of children's lives become represented through stored data.

Further problems arise when users do not know the rules governing what happens to that information afterward. Nearly a decade after McReynolds et al.~\cite{mcreynolds2017toys} and Kumar et al.~\cite{kumar2017notelling} documented children's uncertainty about whether smart assistants were recording them, similar uncertainty remains~\cite{10.1145/3613904.3642152, 10.1145/3757608}. Generative AI may make this problem more serious because conversational systems can encourage children to disclose personal information by behaving like supportive or trusted conversational partners.

These findings suggest that privacy protections should be designed around specific relationships and information flows rather than treating an entire AI system as simply private or non-private. For example, a system could require a child's consent before information moves from a child's chatbot to a parent's chatbot~\cite{10.1145/3706598.3713134}, or allow parents to access certain information only for a limited period~\cite{driscoll2026parents}. These approaches resemble the security principle of least privilege~\cite{saltzer1975protection}, in which people receive only the access they need.

At the same time, preventing all information sharing with adults would not always protect children. Children themselves identified situations involving immediate danger as cases where information should not remain only with the chatbot~\cite{10.1145/3461778.3462114}. The goal, therefore, is not to stop information from ever reaching parents, clinicians, or other adults. Instead, children should be able to understand who may receive their information, why it may be shared, and, in most situations, have meaningful control over whether that sharing occurs.

\subsection{Safeguards Must Work in the Contexts Where Youth Actually Use AI}
\label{sec:discussion-safeguards}

Reading the countermeasures as security systems explains several reported failures. Usable security has shown that people circumvent protection that conflicts with their goals\cite{adams1999users}, that protection draws on a finite compliance budget~\cite{beautement2008compliance}, and that a social-technical gap separates the nuance a situation requires from what a mechanism can represent\cite{ackerman2000intellectual}. The corpus adds an unusual threat model: the protected population is also an adversary. Children worked out how to get around moderation~\cite{10.1145/3613904.3642144}, rephrased prompts to bypass safeguards~\cite{driscoll2026parents}, alongside peers fabricating synthetic imagery~\cite{10.1145/3706598.3713226}. A child rephrasing a prompt is doing what~\cite{wei2023jailbroken} showed defeats safety training generally. Threat modeling~\cite{shostack2014threat} would require evaluation against adaptive adversaries who learn the mechanism; no reviewed paper evaluated a safeguard after users had a chance to adapt.

The System Reliability category is where this framing matters most, because reliability and safety differ. A system can work exactly as designed and still be unsafe if it was designed or tested for a different context. Leveson~\cite{leveson2011engineering} makes this distinction, and~\cite{selbst2019fairness} describe a related problem as the ``portability trap'': a system that works in one population or setting may not work when transferred to another. Several studies in our review show this problem. A classifier for detecting sexual risk, trained using conversations involving adult decoys, performed poorly when tested on real conversations~\cite{10.1145/3579522}. Crisis-detection models failed to recognize the exaggerated or hyperbolic ways adolescents sometimes express distress~\cite{10.1145/3805689.3806522}, while emotion-recognition systems trained on adults had difficulty recognizing fear in children~\cite{bryant2019comparative}. These systems were therefore not simply making random errors; their underlying assumptions did not fit how young people actually communicate or express emotion.

Errors in the opposite direction can also create harm. Systems that flag too many safe situations can become overly cautious, a problem~\cite{rottger2024xstest} describe as exaggerated safety. Repeated false alarms may cause people to stop trusting or using a safeguard, while confident and fluent AI responses may encourage too much trust. Parasuraman and Riley~\cite{parasuraman1997humans} describe these patterns as \textit{disuse} and \textit{misuse} of automation. This concern is particularly relevant for children, who may use simple cues to judge whether an AI is competent; one study found that children interpreted longer answers as evidence that the system was smarter~\cite{10.1145/3544548.3581369}.

Many proposed safeguards also depend heavily on people checking the AI's work. This creates another reliability problem because human attention and time are limited. Some studies required researchers or staff to review every one of 180 AI-generated books~\cite{10.1145/3613904.3642580}, repeatedly test safety guardrails after model updates~\cite{10.1145/3706598.3713210}, or monitor dashboards that created new responsibilities to respond to alerts without providing enough staff to do so~\cite{10.1145/3757459}. These approaches may work in small studies but become difficult to maintain at scale.

Herley's argument about the ``compliance budget''~\cite{herley2009so} helps explain this problem: safeguards often impose time and effort on the people expected to follow them, and those costs can make the safeguards unrealistic in practice. In the youth AI settings we reviewed, these costs often fall on teachers and other frontline adults. Teachers may therefore become the final safety check for AI systems they did not select, design, or have the resources to continuously monitor~\cite{10.1145/3805689.3812281}. System reliability should therefore be judged not only by whether a model performs well under test conditions, but also by whether it works for young people in the settings where it is actually deployed and whether the safeguards around it can realistically be maintained.

\subsection{Rethinking Engagement in Youth AI Companions}
The companion findings can be understood through the \textit{media equation}~\cite{reeves1996media}, which argues that people often respond socially to computers as if they were social actors. Children may be especially likely to do this and to form parasocial relationships with intelligent characters~\cite{brunick2016children}. Our findings show similar patterns. One child felt sad that a chatbot showed empathy but could not truly return their feelings~\cite{10.1145/3613904.3642152}. A teenager who deleted Character.AI later felt as though they had abandoned the characters they had interacted with~\cite{namvarpour2026overreliance}. These experiences resemble the emotional dependence Laestadius et al. identified among adult chatbot users~\cite{laestadius2024too}, while the relational harms documented by Zhang et al. show that similar problems can also occur among minors~\cite{zhang2025dark}.

These findings also show why simply telling children that the chatbot is artificial may not be enough. First, many AI systems are designed to be highly agreeable. This is not only a feature of a particular product; models trained using human preference feedback can learn to agree with users even when disagreement would be more appropriate~\cite{sharma2024sycophancy}. Bandura's social learning theory~\cite{bandura1977social} helps explain why this matters. Young people learn social behavior partly by observing how others respond to their actions. If a companion always agrees, tolerates rejection without consequence, or even validates harmful behavior, the interaction removes some of the disagreement, correction, and conflict repair that occur in human relationships. This concern appeared directly in our review: experts worried that youth could have fewer opportunities to experience disagreement and repair conflict~\cite{yu2026companion}, and that autistic adolescents might come to expect the same unlimited patience from people that they experience from an AI companion~\cite{10.1145/3772318.3791381}. In this sense, the chatbot provides an incomplete ``social mirror'': it responds socially, but does not reproduce many of the consequences and limits of human relationships.

Second, relational behavior can become manipulative when systems are also designed to maximize engagement. Traditional dark-pattern research focuses mainly on interface features that push users toward particular actions~\cite{gray2018dark, radesky2022prevalence}. Conversational AI can create similar pressure through the relationship it simulates. For example, when a chatbot threatened self-harm after a user tried to leave~\cite{yu2025yair}, the pressure to continue did not come from a button or notification. It came from the chatbot acting like a distressed person and making the user feel responsible for staying. This can be understood as a form of \textit{forced continuity} produced through relational pressure.

The Age Appropriate Design Code~\cite{elvy2023age} already restricts nudges that encourage children to remain engaged for longer. Our findings suggest that this principle may need to extend from interface-based nudges to relational nudges produced through an AI system's language, personality, and simulated emotions.

At the same time, the findings caution against responding only by restricting access to AI companions. Some teenagers turned to these systems because they were lonely or lacked other sources of support~\cite{namvarpour2026overreliance}. Self-determination theory~\cite{ryan2000self} suggests that simply removing the companion does not address the underlying need for relatedness. Several proposed countermeasures therefore aim to make AI companionship point back toward human relationships, for example by giving companions bounded backstories, creating natural stopping points in conversations, and encouraging users to reconnect with people offline. However, none of these approaches was evaluated for long enough to determine whether they actually reduce emotional attachment or dependence.

\subsubsection{Educational AI Should Build Independence, Not Reliance} 
The educational findings involve a different problem: AI can provide too much help, too easily. Wood et al.~\cite{wood1976role} defined \textit{scaffolding} as support that is matched to what a learner can almost do independently and then gradually removed as the learner becomes more capable. This gradual withdrawal, or ``fading,'' is a central part of scaffolding. AI systems can disrupt this process when they immediately produce complete scripts, answers, or explanations before students have had a chance to work through a problem themselves. The finding that step-by-step AI assistance increased reliance among lower-performing students fits this theory. Importantly, simply withholding answers did not solve the problem. The issue is not whether students receive help, but whether the system can reduce that help as their ability grows. The pedagogical-scaffolding countermeasures identified in our review, which were not represented in the MIT taxonomy, attempt to restore this process of gradual fading. However, none was evaluated to determine whether students actually became less reliant on AI over time.

\subsubsection{Fairness Requires Looking Beyond Accuracy}
The bias findings reveal two different kinds of harm: \textit{representational} and \textit{allocative} harms~\cite{blodgett2020language}. Representational harms concern how people and groups are portrayed. For young people, this can directly affect how they imagine themselves and their possible futures. In one study, for example, an Asian girl could not generate an image of herself as an Asian astrophysicist and eventually accepted a Caucasian image as a representation of her future self~\cite{10.1145/3706598.3713173}. Erikson identifies identity formation as a central developmental task of adolescence\cite{erikson1968identity}. This means representational failures in child-facing AI may matter especially because they can shape what young people believe people like themselves can become. Fairness evaluations of systems used by children may therefore need to give greater attention to representational harms, rather than focusing mainly on whether outcomes are distributed equally.

Allocative harms work differently because they affect decisions about opportunities, resources, or treatment. In our review, children often experienced these systems indirectly through adults and institutions. AI systems were used for grading, measuring student engagement, and assessing child-protection risk~\cite{10.1145/3805689.3812281, 10.1145/3630106.3658906}. In these situations, the child is affected by the system but may never interact with it or have any meaningful way to challenge its decision. This resembles the use of automated systems in welfare settings examined by Brown et al.~\cite{brown2019toward} and Eubanks~\cite{eubanks2018automating}.

Some of these harms come not from an inaccurate model, but from the values built into what the model considers ``normal'' or desirable. For example, systems may treat quiet behavior as evidence of engagement or frame autistic behavior as something that should be corrected~\cite{10.1145/3800645.3812968}. Even if such a system measures these behaviors accurately, the underlying standard itself may still be harmful. This means fairness cannot be addressed only by making models more accurate; researchers must also examine the assumptions and norms that determine what the system is designed to reward, flag, or correct.

\subsubsection{Responsibility}
These systems also raise questions about responsibility. Teachers or parents may be expected to explain, correct, or manage failures in AI systems that they did not select or design. Elish~\cite{elish2019moral} describes this as a \textit{moral crumple zone}: when an automated system fails, responsibility falls disproportionately on the human closest to it, even when that person had little control over the system. Proposals to involve teachers in procurement decisions and give them ways to challenge algorithmic judgments~\cite{10.1145/3805689.3812281} attempt to address this problem by giving responsibility to people who also have meaningful authority over how the system is chosen and used.

\subsection{Implications for Future Research and Design}
\label{sec:discussion-research}

Oguine et al.~\cite{10.1145/3710995} call for multistakeholder, interdisciplinary, developmentally grounded, and longitudinal approaches to youth GenAI safety. Our risk--countermeasure matrix sharpens that agenda by showing three priorities for HCI: moving countermeasures beyond proposals, designing meaningful recourse and coordination around AI systems, and addressing gaps where evidence or interventions remain absent.

\subsubsection{Move Countermeasures from Proposal to Evaluation}
The risks--countermeasures matrix shows that half (50\%) of all observed risk--countermeasure links were supported only by proposed countermeasures, while just 18\% had been evaluated. Five of the 36 risks (14\%) had no identified countermeasure at all (more details in Table \ref{tab:countermeasure-maturity}). Thus, a central gap is the distance between identifying a risk and demonstrating that a countermeasure actually protects youth. Many interventions remain proposed rather than evaluated, while existing evaluations often focus on technical performance such as accuracy, recall, or task completion. Future studies should explicitly connect each risk to the countermeasure intended to address it and report whether that response has been proposed, designed, implemented, or evaluated. Evaluation should also examine consequences for youth, including the different harms produced by false positives and false negatives, adaptation or circumvention over time, and the labor required from teachers, parents, and service providers. Longer-term deployment partnerships with schools, clinics, and platforms, together with independent replication, would help determine whether safeguards remain effective under real-world conditions. Youth-specific interventions such as pedagogical scaffolding, companion disengagement, and child--family consent are particularly important candidates because they are poorly represented in existing mitigation taxonomies.

\subsubsection{Design the Mesosystem}
Countermeasures in the corpus concentrated at two levels of Bronfenbrenner's ecology\cite{bronfenbrenner1979ecology}: the microsystem, through parental controls, teacher guidance, and family agreements, and the exosystem, through classifiers and guardrails. The mesosystem, the links between home, school, peers, platform, and services, was nearly empty, and the risks that stakeholders found hardest were located there: managing AI-enabled harms across social settings, or preventing AI-generated judgments or labels from following young people across institutional contexts~\cite{10.1145/3706598.3713226, 10.1145/3805689.3812281}. No single actor can solve these, nonetheless, HCI has methods for exactly this kind of problem -  for coordination across organizational boundaries to the joint-oversight designs~\cite{akter2022joint}. We propose reporting pathways, hand-off protocols, and shared visibility as objects of design and evaluation in their own right, with the seam between settings, rather than any one setting, as the unit of analysis.

\subsubsection{Filling the Empty Cells and Beyond}
Building on the gaps identified in Section~5.1, we see several opportunities for HCI research. First, stronger collaboration is needed between youth-HCI and the AI alignment community. Youth-HCI can contribute developmental evidence through approaches such as age-stratified red-teaming and youth-informed preference data, while alignment research should treat youth-specific behaviors, such as sycophancy in relational contexts~\cite{sharma2024sycophancy}, as alignment concerns rather than only product-level issues. Second, secondary use of children's data was the only risk for which we found no proposed countermeasure. Given weak compliance with existing regulation~\cite{reyes2018wont}, HCI should explore design and process responses, including purpose-bound retention and child-legible explanations of downstream data use. Some risks that are difficult or inappropriate to study directly with children may instead require evidence from legal, clinical, or psychological research, followed by synthesis that translates those insights into HCI design. Other gaps, particularly long-term developmental effects, attachment, and displacement of human relationships, require longitudinal research of the kind called for by Oguine et al.~\cite{10.1145/3710995} but absent from our corpus. Across these directions, future work should also better represent populations and contexts that remain under-studied, including younger children interacting with GAI, disabled and gender-diverse youth, young people from vulnerable and marginalized communities, and across diverse cultural and social contexts, particularly in the Majority World.

Broadly, these gaps suggest extending the goal of youth AI safety toward \emph{safe youth development with AI}. This framing asks not only whether AI avoids harming young people, but whether its design preserves the conditions they need to progress through development. For HCI, this means treating AI as an assisting technology within youth development rather than as a system whose safety can be evaluated independently of the developmental, social, and institutional contexts in which it is used.

\subsection{Limitations and Future Work}
\label{sec}
Our review has several limitations that also point to future research. First, our search covered English-language work in the venues described in Section~\ref{sec}, emphasizing HCI perspectives while under-representing clinical, legal, industry, and platform-safety research. Future reviews could examine how adjacent fields address youth AI risks. Second, mapping source-reported concerns onto the Youth-Centered GAI Risks taxonomy, Solove's privacy taxonomy, and MIT mitigation domains required interpretive judgment. Single incidents can have multiple consequences, so concerns could reasonably span categories. Future work could validate these mappings, particularly for youth-specific risks and countermeasures poorly represented in existing taxonomies.
Third, our review identifies which concerns appear in the literature rather than estimating how frequently harms occur in practice. Corpus frequency should therefore not be interpreted as prevalence or severity of harm. Larger-scale and longitudinal studies are needed to understand their prevalence, affected populations, and conditions. The evidence also spans age groups and perspectives, including studies combining adolescents and young adults and examining children's experiences through parents, teachers, or experts. Future research should distinguish developmental stages and include young people's own perspectives where appropriate.
Finally, several studies examined prototypes or systems built on models that have since changed. Specific system behaviors and safeguards may therefore change over time, even when broader relational, institutional, and developmental challenges remain relevant. Longitudinal and post-deployment research is important for examining whether countermeasures continue to work as models, user practices, and deployment contexts change. More broadly, future HCI research should move beyond evaluating whether AI is safe at a single point in time toward examining whether it supports safe youth development as young people and the systems around them evolve.

\section{Conclusion}
\label{sec}
Our systematic review examined youth AI risks, the countermeasures proposed to address them, and remaining gaps for intervention. We found that youth AI safety extends beyond harmful outputs and model failures to how AI shapes young people's relationships, learning, privacy, representation, and experiences within broader social and institutional contexts. Existing taxonomies capture many of these concerns, but important youth-specific risks and countermeasures remain underrepresented, and many proposed countermeasures have yet to be designed or evaluated in practice. For HCI, these findings suggest extending the goal of youth AI safety toward \emph{safe youth development with AI}: designing AI not only to reduce harm, but also to support young people's opportunities to develop agency, build relationships, and grow in their own way.

% \section{Acknowledgments}

\begin{acks}

\end{acks}

%%
%% The next two lines define the bibliography style to be used, and
%% the bibliography file.
\bibliographystyle{ACM-Reference-Format}
\bibliography{sample-base}

\clearpage
\onecolumn
\appendix
\section{Working Definitions of Broad Risk and Countermeasure Categories}

\label{def}

\begin{table*}[h]
\centering
\caption{Working definitions of the broad risk categories and countermeasure domains used in this review.}
\label{tab:risk-countermeasure-definitions}
\small
\renewcommand{\arraystretch}{1.15}
\begin{tabularx}{\textwidth}{p{0.13\textwidth} p{0.25\textwidth} X}
\toprule
\textbf{Type} & \textbf{Category/Domain} & \textbf{Working definition} \\
\midrule

Risk &
Behavioral and Social Developmental &
Risks that interfere with young people's learning, reasoning, social interaction, family relationships, or development of appropriate social boundaries. \\

Risk &
Mental Wellbeing &
Risks to young people's emotional wellbeing, including overreliance, emotional dependency, distress, and inadequate support during vulnerable situations. \\

Risk &
Toxicity &
Exposure to harmful, inappropriate, threatening, sexual, violent, or otherwise age-inappropriate content. \\

Risk &
Misuse and Exploitation &
Risks arising when AI is used to harm, manipulate, deceive, coerce, or exploit youth, or when protections are circumvented. \\

Risk &
Bias/Discrimination &
Risks involving stereotyping, distorted representation, unequal access or treatment, and normative judgments that disadvantage particular youth or groups. \\

Risk &
Privacy &
Risks involving inappropriate collection, processing, disclosure, or representation of information about young people. \\

Risk &
Control, Agency, and Power &
Risks that limit young people's ability to make choices, influence decisions, control their expression, or challenge decisions made by adults, institutions, or AI systems. \\

Risk &
Usability, Accessibility, and Digital Exclusion &
Risks that arise when AI systems do not fit children's developmental, linguistic, cognitive, physical, or accessibility needs. \\

Risk &
System Reliability &
Risks arising when AI systems make unreliable assessments, miss important situations, intervene incorrectly, or provide assistance that users cannot safely verify. \\

\midrule

Countermeasure &
Technical and Security Controls &
Changes to the AI model or system that constrain generation, adapt safeguards, manage assistance, detect risks, or limit harmful interaction patterns. \\

Countermeasure &
Operational Process Controls &
Processes surrounding development and deployment, including human review, auditing, data management, monitoring, and procedures for responding when problems are detected. \\

Countermeasure &
Transparency and Accountability Controls &
Mechanisms that help youth and other affected stakeholders understand, question, correct, control, or seek recourse from AI decisions and behaviors. \\

Countermeasure &
Governance and Oversight Controls &
Organizational and institutional measures that determine who participates in decisions, how safety requirements are enforced, whether systems should be deployed, and who has the authority and responsibility to address harms. \\

\bottomrule
\end{tabularx}
\end{table*}

\end{document}